\documentclass[twocolumn,english]{revtex4-1}
\usepackage[T1]{fontenc}
\usepackage[latin9]{inputenc}
\usepackage{amsmath}
\usepackage{amssymb}
\usepackage{graphicx}
\usepackage{esint}
\usepackage{nameref}
\usepackage{subfigure}

\makeatletter
\usepackage{babel}

\makeatother

\usepackage{babel}
\begin{document}

\title{Exact coherence dynamics mediated by a single cavity mode in the presence of leakage}

\author{Yusui Chen$^{1,2}$}
\email{ychen132@nyit.edu}
\author{Quanzhen Ding$^{2}$}
\author{Wufu Shi$^{2,4}$}
\author{Jun Jing$^{3}$}
\email{jingjun@zju.edu.cn}
\author{Ting Yu$^{2}$}

\affiliation{$^{1}$ Department of Physics, New York Institute of Technology, Old Westbury, NY 11568, USA ~\\
$^{2}$ Department of Physics, Stevens Institute of Technology, Hoboken, NJ 07030, USA~\\
$^{3}$ Department of Physics, Zhejiang University, Hangzhou, 310027, Zhejiang, China\\
$^{4}$ Beijing Computational Science Research Center, Beijing 100084, China}

\date{\today}
\begin{abstract}
We present a quantum-state-diffusion equation to characterize the dynamics of a generic atomic system coupled to a leaky cavity mode. As quantum resources, the population, the coherence and even the entanglement of the system would gradually leak out of the cavity. The effect from the leakage of the cavity-mode to the uncontrollable degree of freedom, e.g., environment, is however not always negative to particular targets. A well-established scenario is that a photon counter attached to the cavity would absorb and then record the leaking photon. As a medium between the system and the photon counter, a strong coupling between system and cavity is necessary to enhance the measurement efficiency. While we find it also reduce the reading efficiency of the photon counter at the same time as fewer photons leak out of the cavity. We investigate the competition between these two mechanisms in the framework of non-Markovian open-quantum-system dynamics, where the photon counter serves as a general bosonic environment. Our results provide an optimized parameter space for system entanglement preservation and generation.
\end{abstract}

\pacs{03.65.Yz, 03.67.Bg, 03.65.Ud, 32.90.+a}
\maketitle

\section{Introduction}

Cavity quantum electrodynamics (CQED) is a field focusing on the coherent light-matter interactions \cite{scullybook,cqed1,cqed2}, which lay the foundation for a wide-range field of quantum optics, including quantum information processing, quantum sensing, quantum simulating and quantum computing \cite{cqed-grange, cqed-nori, cqed-aspelmeyer, cqed-Kippenberg, cqed-JKThompson, cqed-tkm, cqed-JPHome, cqed-sank}. The Jaynes-Cummings (JC) Hamiltonian arising naturally for CQED systems has been well studied to understand energy exchange and state shift between a single qubit and the coupled harmonic oscillator. Similarly, extended Jaynes-Cummings models focus on interplays between the multilevel atoms or a collection of spin-$1/2$ particles and the quantized electromagnetic field. Heterogeneous optical systems are engineered to enhance the coherent light-matter interactions, to precisely tune spin-spin interactions using the driven optical cavity, and to exceed the Heisenberg limit in super-sensitive quantum measurements \cite{usc-leroux, cqed-pp,cqed-hamsen, cqed-kohler,cqed-ychen}.

A challenge facing these goals is to understand the temporal evolution of qubits or artificial atomic spins in CQED systems. The difficulties arise when the optical cavity is imperfect and has a chance to lose photons, which is crucial and inevitable in a real experimental setup. The fact that a cavity loses photon due to couplings to the environment, the readout detectors or photon counters, namely the cavity loss, indicates that the dynamics of the JC model should be investigated in the quantum open-system formalism \cite{breuerbook, zollerbook, nm-vega, nm-HPB}. However, there are limited exact approaches to investigate the evolution of the atoms trapped in the leaky cavity \cite{tool-hu, tool-paz, tool-goan, tool-zhang}. Conventionally, we assume that these systems of interest are sufficiently isolated from the environment. As a result, the weak coupling approximation and the Markov approximation are valid. Therefore, the impact of the environment is considered as a perturbation. However, the weak-coupling approximation is not always applicable. For example, when the cavity is attached to a readout detector, it requires strong couplings between the qubits and the cavity mode, in order to realize high-frequent measurements on the qubit system. And in some realistic systems, e.g., the optomechanical system, the external noise naturally is non-Markovian, rather than Markovian \cite{cqed-aspelmeyer}.

One motivation of this paper is to develop a systematic and exact approach to study the dynamics of the trapped atoms, from the Markov limit to the non-Markovian situation. The quantum-state-diffusion (QSD) approach provides a simple basis to obtain the reduced density matrices of the systems of interest, by taking the ensemble average over all the possible trajectories, each described by a stochastic Schr\"{o}dinger equation \cite{qsd-gisin, qsd-strunz, qsd-jing}. So far, the QSD approach has successfully resolved the non-Markovian dynamics for the continuous variable systems, e.g., optical cavities and mechanical oscillators, and the discrete variable systems, e.g., the multiple qubits and the multilevel systems, in the presence of non-Markovian environments \cite{strunz2018, strunz2014, jing2018, jing2015, shi2018, chen2017,chen2014,jing2013, zhao2011,ma2014,shi2013}. To investigate the embedded atomic systems, we model the single cavity mode and the surrounding environment serving as the attached photon detector as two cascaded environments, then the trajectories of the embedded systems can be expanded in the basis of two independent Gaussian random processes, and the reduced density matrices can be recovered by averaging over these two processes.

Another motivation is that the derived QSD equations can be used to numerically study the non-Markovian features induced by the environment. Then it allows one to optimize parameters, particularly the coupling strength between the atom and the cavity mode, and that between the cavity mode and its environment, to enhance the state transition fidelity and efficiency and reduce the photon loss due to the dissipative environment. Although the dissipative environment usually is harmful to the system coherence, it is shown that in a particular parameter regime, the quantum entanglement could arise due to the coupling between cavity and environment. Besides, the QSD equations can be used to modulate the steady state of the embedded system or enhance its coherence by engineering the external environment.

The paper is organized as following: In Sec. II, we briefly review the QSD method and the standard non-Markovian QSD approach and introduce a general ansatz for the two cascaded environments model. In Sec.III, we derive the QSD equation for a two-qubit system. We also optimize the coupling strength between the two layers of environments. Some unique non-Markovian phenomena are discussed based on the numerical simulations. The paper is concluded in Sec. IV.

\section{Generic Models and the Quantum-state Diffusion Approach}

For a generic quantum open system model, the total Hamiltonian in the interaction picture is (setting $\hbar=1)$
\begin{equation}
H_{{\rm tot}}=H_{{\rm sys}}+L\sum_{k}g_{k}^{*}b_{k}^{\dagger}e^{i\omega_{k}t}+L^{\dagger}\sum_{k}g_{k}b_{k}e^{-i\omega_{k}t},
\end{equation}
where $L$ is the Lindblad operator of system and $b_{k}(b_{k}^{\dagger})$ is the annihilation (creation) operator of $k$th mode of the bosonic environment, and the corresponding complex coupling strength is $g_k$. Since the system and the environment as a whole is closed, their dynamics can be characterized by Schr\"{o}dinger equation for the closed system, $\partial_t |\Psi\rangle = -iH_{\rm{tot}}|\Psi\rangle$, where $|\Psi\rangle$ is the total quantum state of the system and the environment. Expanding the Schro\"{o}dinger equation in the basis of the Bargmann coherent state, $b_{k}|z\rangle=z_{k}|z\rangle$, we obtain a stochastic Schr\"{o}dinger equation for the system of interest,
\begin{equation}
\partial_{t}\psi_{t}(z^{*})=\left[-iH_{{\rm sys}}+Lz_{t}^{*}-L^{\dagger}\bar{O}(t,z^{*})\right]\psi_{t}(z^{*}), \label{eq:original_qsd}
\end{equation}
where $\psi_{t}(z^{*})=\langle z|\Psi\rangle$ is the quantum trajectory and $\langle z|=\langle z_{1},z_{2},...,z_{k},...|$. $z_{t}^{*}=-i\sum_{k}g_{k}^{*}z_{k}^{*}e^{i\omega_{k}t}$ is a Gaussian random process. For the zero-temperature environment, the correlation function of the noise $z_t^*$ is $\alpha(t,s)=\sum_{k}|g_{k}|^{2}e^{-i\omega_{k}(t-s)}$. In the last term of Eq.~(\ref{eq:original_qsd}), the operator $\bar{O}(t,z^*)=\int_{0}^{t}ds\alpha(t,s)O(t,s,z^{*})$, where the operator $O(t,s,z^*)$ is defined as $\frac{\delta\psi_t}{\delta z_s^*}=O(t,s,z^*)\psi_t$, and governed by a nonlinear differential equation,
\begin{equation}
\frac{\partial}{\partial t}O=[-iH_{{\rm sys}}+Lz_{t}^{*}-L^{\dagger}\bar{O},O]-L^{\dagger}\frac{\delta\bar{O}}{\delta z_{s}^{*}}.\label{eq:O_consistency}
\end{equation}

In this paper, we consider a leaky cavity mode, and its total Hamiltonian $H_{\rm{tot}}$ can be written as
\begin{equation}
H_{{\rm tot}}=H_s+H_{s-c}+H_{c}+H_{c-e}+H_e,\label{eq:total_H}
\end{equation}
where
\begin{align*}
H_c & =\sum_{k}\omega_{k}a_{k}^{\dagger}a_{k},\\
H_e & =\sum_{k'}\omega_{k'}b_{k'}^{\dagger}b_{k'},\\
H_{s-c} & =\sum_{k}(g_{k}^{*}La_{k}^{\dagger}+g_{k}L^{\dagger}a_{k}),\\
H_{c-e} & =\sum_{k,k'}(g_{k}^{*}f_{k'}^{*}a_{k}b_{k'}^{\dagger}+g_{k}f_{k'}a_{k}^{\dagger}b_{k'}).
\end{align*}
The total Hamiltonian consists of the atomic system $H_s$, the leaky cavity modes $H_c$, the surrounded dissipative environment $H_e$, the interaction between system and cavity $H_{s-c}$ and the coupling between cavity and environment $H_{c-e}$.

\begin{figure}[h]
\includegraphics[scale=0.5]{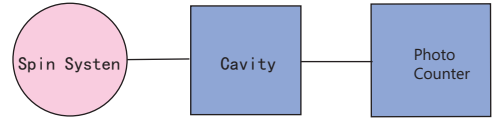}

\caption{Schematic of the qubit system measured by a cavity quantum
probe which is coupled to a photon counter. }
\end{figure}

To apply the QSD approach, we expand the cavity mode and the surrounding zero-temperature bosonic environment independently, in the basis of the Bargmann coherent states, $|z\rangle$ and $|y\rangle$, respectively, satisfying $\langle z|a_{k}^{\dagger}=z_{k}^{*}\langle z|$ and $\langle y|b_{k}^{\dagger}=y_{k}^{*}\langle y|$. Hence, the formal QSD equation can be written as
\begin{align}
\partial_{t}\psi_{t} & =\Biggl[-iH_{s}+Lz_{t}^{*}-L^{\dagger}\int_{0}^{t}{\rm d}s\alpha(t,s)O_{z}\nonumber \\
 & -iy_{t}^{*}\int_{0}^{t}{\rm d}s\alpha(t,s)O_{z}-iz_{t}^{*}\int_{0}^{t}{\rm d}s\beta(t,s)O_{y}\Biggl]\psi_{t},\label{eq:qsd_general}
\end{align}
where $\psi_{t}(z^{*},y^{*})=\langle z,y|\Psi\rangle$ denotes the quantum state trajectory consisting of two Gaussian noises, $z_t^*$ and $y_t^*$, defined as
\begin{align*}
z_{t}^{*} & =-i\sum_{k}g_{k}^{*}z_{k}^{*}e^{i\omega_{k}t},\\
y_{t}^{*} & =-i\sum_{k'}f_{k'}^{*}y_{k'}^{*}e^{i\omega_{k'}t}.
\end{align*}
The correlation functions for the two noises are
\begin{align*}
\alpha(t,s) & =\sum_{k}|g_{k}|^{2}e^{-i\omega_{k}(t-s)},\\
\beta(t,s) & =\sum_{k'}|f_{k'}|^{2}e^{-i\omega_{k'}(t-s)}.
\end{align*}
Two noises-dependent operators $O_{z}(t,s,z^*, y^*)$ and $O_{y}(t,s,z^*, y^*)$ are defined as
\begin{align}
\frac{\delta\psi_{t}}{\delta z_{s}^{*}} & =O_{z}(t,s,z^{*},y^{*})\psi_{t},\\
\frac{\delta\psi_{t}}{\delta y_{s}^{*}} & =O_{y}(t,s,z^{*},y^{*})\psi_{t},
\end{align}
and governed by the following two equations,
\begin{align}
\partial_{t}O_{z} & =\left[-iH_{eff},\,O_{z}\right]-L^{\dagger}\frac{\delta\bar{O}_{z}}{\delta z_{s}^{*}}-iy_{t}^{*}\frac{\delta\bar{O}_{z}}{\delta z_{s}^{*}}-iz_{t}^{*}\frac{\delta\bar{O}_{y}}{\delta z_{s}^{*}},\label{eq:consist1}\\
\partial_{t}O_{y} & =\left[-iH_{eff},\,O_{y}\right]-L^{\dagger}\frac{\delta\bar{O}_{z}}{\delta y_{s}^{*}}-iy_{t}^{*}\frac{\delta\bar{O}_{z}}{\delta y_{s}^{*}}-iz_{t}^{*}\frac{\delta\bar{O}_{y}}{\delta y_{s}^{*}},\label{eq:consist2}
\end{align}
where the effective Hamiltonian $H_{eff}$ is
\begin{equation}
H_{eff}=H_{s}+iLz_{t}^{*}-iL^{\dagger}\bar{O}_{z}+y_{t}^{*}\bar{O}_{z}+z_{t}^{*}\bar{O}_{y},
\end{equation}
 where the operators $\bar{O}_z$ and $\bar{O}_y$ are respectively defined as $\bar{O}_{z}=\int_{0}^{t}ds\alpha(t,s)O_{z}$ and $\bar{O}_{y}=\int_{0}^{t}ds\beta(t,s)O_{y}$. In the QSD approach, the basis of the $O$ operator is dependent on the system Hamiltonian $H_{sys}$ and lives only in the Hilbert space of the system. The coefficient functions are determined by the evolution equations. The basis for the O operators in this two-layer noise scenario is found to be the same as that in the single-layer noise. The coupling between cascaded environments does not change the basis of the $O$ operator, but modify the evolution of the coefficient functions. In addition, the $O$ operator of the leaky cavity case will reduce to the $O$ operator in Eq. (\ref{eq:original_qsd}), as the limit when the memory time parameter of the leaky mode is zero. By considering the above mentioned requirements, we derive the two $O$ operators in the form of
\begin{align}
O_{z}(t,s,z^{*}) & =O(t,s)-i\int_{0}^{s}{\rm d}\tau\beta(t,\tau)O_{y}(t,\tau,z^{*}),\nonumber \\
O_{y}(t,s,z^{*}) & =-i\int_{0}^{s}{\rm d}\tau\alpha(t,\tau)O_{z}(t,\tau,z^{*}).\label{eq:new_ansatz}
\end{align}
The $O(t,s)$ in the ansatz is the original $O$ operator in the regular JC model. It is easy to verify that this $O$ operator consists of that in a regular JC model, when the correlation function $\beta(t,s)=0$. Notably, the upper bound of the integration is the time index $s$, not $t$. This indicates that the non-Markovian dynamics emerges as a compromise result of the cavity mode and the external environment. When both correlation functions $\alpha(t,s)$ and $\beta(t,s)$ take the form of a Dirac delta function, the Markov limit as a boundary condition has been verified, $O_{y}=0$ and $O_{z}=L$. Actually, our discussion can be extended to an arbitrary size of linear-coupled cavity-network.

\section{QSD equation for the two-qubit system coupled to a leaky cavity mode}

As the building block of quantum information and quantum computation, the coupled two-qubit model is popular in studying the quantum entanglement, quantum decoherence process and various quantum features. In this section, we focus on the coupled two-qubit system in the presence of a leaky cavity mode. The formal total Hamiltonian (\ref{eq:total_H}) is
\begin{equation}
H_{tot}=H_{s}+H_{s-c}+H_{c}+H_{c-e}+H_{e},
\end{equation}
where
\begin{align*}
H_{s} & =\frac{\omega_{s}}{2}\left(\sigma_{z}^{A}+\sigma_{z}^{B}\right),\\
H_{c} & =\omega a^{\dagger}a,\\
H_{e} & =\sum_{k}\omega_{k}b_{k}^{\dagger}b_{k},\\
H_{s-c} & =g^{*}La^{\dagger}+gL^{\dagger}a,\\
H_{c-e} & =a\sum_{k}f_{k}^{*}b_{k}^{\dagger}+a^{\dagger}\sum_{k}f_{k}b_{k}.
\end{align*}
$L=\kappa_{1}\sigma_{-}^{A}+\kappa_{2}\sigma_{-}^{B}$ is the coupling operator in the Hilbert space of the system. The operator $a$ in the Hilbert space of the cavity is the linear coupling operator between the cavity and the dissipative environment. Then, the formal QSD equation for this model is given as
\begin{equation}
\partial_t\psi_t = \left[-i H_s + Lz_t^* -(L^{\dagger}+iy_t^* )\bar{O}_z - iz_t^* \bar{O}_y\right]\psi_t. \nonumber
\end{equation}
Particularly, for a single cavity mode, the noise $z_t^* = -ig^{*}z^*e^{i\omega t}$ has an explicit correlation function $\alpha(t,s) = |g|^2e^{-i\omega(t-s)}$. The second noise $y_t^* = -i\sum_k \frac{f_k^*}{g}y_k^*e^{i\omega_k t}$ is characterized by the correlation function $\beta(t,s) = \frac{1}{|g|^2}\sum_k |f_k|^{2}e^{-i\omega_k (t-s)}$. The factor $1/|g|^2$ in the correlation function $\beta(t,s)$ indicates the competition between the controlling process and the readout process. When the parameter $g$ is large enough, one can control the embedded system via the cavity mode. However, the external readout detector cannot receive any signal with sufficient strength. On the contrary, a smaller parameter $g$ can increase the readout probability, but decrease the coupling between the qubit system and the cavity mode. We will show some optimized results for the effect of $g$ on the system coherence (entanglement) in the next section.

For the two-qubit system in a dissipative environment, the $O$ operator can be expanded as
\begin{widetext}
\begin{align}
O & =\sum_{j=1}^{4}f_{j}(t,s)O_{j}+i\int_{0}^{t}{\rm d}s'f_{5}(t,s,s')z_{s'}^{*}O_{5},\label{eq:2spin_original_O}
\end{align}
where the five operators are explicitly given as
\begin{align*}
O_{1} & =\sigma_{-}^{A},\,O_{2}=\sigma_{-}^{B},\ O_{3}=\sigma_{z}^{A}\sigma_{-}^{B}, O_{4}=\sigma_{-}^{A}\sigma_{z}^{B},\,O_{5}=\sigma_{-}^{A}\sigma_{-}^{B}.
\end{align*}
Applying the ansatz (\ref{eq:new_ansatz}), the two operators $O_z$ and $O_y$ are explicitly written in the form of
\begin{align*}
O_{z} & =\sum_{j=1}^{4}n_{j}(t,s)O_{j} +i\int_{0}^{t}{\rm d}s'\left(n_{5}(t,s,s')z_{s'}^{*}+n_{6}(t,s,s')y_{s}^{*}\right)O_{5}, \\
O_{y} & =\sum_{j=1}^{4}m_{j}(t,s)O_{j}\label{eq:2spin_O}+i\int_{0}^{t}{\rm d}s'\left(m_{5}(t,s,s')z_{s'}^{*}+m_{6}(t,s,s')y_{s}^{*}\right)O_{5}.
\end{align*}
Substituting the above ansztz into Eqs.~(\ref{eq:consist1}) and (\ref{eq:consist2}), all of the coefficient functions are found to be governed by a group of partial differential equations
\begin{align*}
\partial_{t}n_{1} & =i\omega_{A}n_{1}+\kappa_{1}N_{1}n_{1}+\kappa_{1}N_{4}n_{4}-\kappa_{2}N_{1}n_{3}+\kappa_{2}N_{3}n_{1}+\kappa_{2}N_{3}n_{4}+\kappa_{2}N_{4}n_{3}-\frac{iN_{5}}{2}\kappa_{2}, \\
\partial_{t}n_{2} & =i\omega_{B}n_{2}-\kappa_{1}N_{2}n_{4}+\kappa_{1}N_{3}n_{4}+\kappa_{1}N_{4}n_{2}+\kappa_{1}N_{4}n_{3}+\kappa_{2}N_{2}n_{2}+\kappa_{2}N_{3}n_{3}-\frac{iN_{5}}{2}\kappa_{1}, \\
\partial_{t}n_{3} & =i\frac{\omega_{B}}{2}n_{3}-\kappa_{1}N_{2}n_{1}+\kappa_{1}N_{3}n_{1}+\kappa_{1}N_{4}n_{2}+\kappa_{1}N_{4}n_{3}+\kappa_{2}N_{2}n_{3}+\kappa_{2}N_{3}n_{2}-\frac{iN_{5}}{2}\kappa_{1}, \\
\partial_{t}n_{4} & =i\frac{\omega_{A}}{2}n_{4}+\kappa_{1}N_{1}n_{4}+\kappa_{1}N_{4}n_{1}-\kappa_{2}N_{1}n_{2}+\kappa_{2}N_{3}n_{1}+\kappa_{2}N_{3}n_{4}+\kappa_{2}N_{4}n_{2}-\frac{iN_{5}}{2}\kappa_{2}, \\
\partial_{t}n_{5} & =i(\omega_{A}+\omega_{B})n_{5}+\kappa_{1}N_{1}n_{5}+\kappa_{1}N_{4}n_{5}+\kappa_{1}N_{5}\left(n_{1}-n_{4}\right)+\kappa_{2}N_{2}n_{5}+\kappa_{2}N_{3}n_{5}+\kappa_{2}N_{5}(n_{2}-n_{3}), \\
\partial_{t}n_{6} & =i(\omega_{A}+\omega_{B})n_{6}+\kappa_{1}N_{1}n_{6}+\kappa_{1}N_{4}n_{6}+\kappa_{1}N_{6}\left(n_{1}-n_{4}\right)+\kappa_{2}N_{2}n_{6}+\kappa_{2}N_{3}n_{6}+\kappa_{2}N_{6}(n_{2}-n_{3}),\\
\partial_{t}m{}_{1} & =i\omega_{A}m_{1}+\kappa_{1}N_{1}m_{1}+\kappa_{1}N_{4}m_{4}-\kappa_{2}N_{1}m_{3}+\kappa_{2}N_{3}m_{1}+\kappa_{2}N_{3}m_{4}+\kappa_{2}N_{4}m_{3}-\frac{iN_{6}}{2}\kappa_{2}, \\
\partial_{t}m_{2} & =i\omega_{B}m_{2}-\kappa_{1}N_{2}m_{4}+\kappa_{1}N_{3}m_{4}+\kappa_{1}N_{4}m_{2}+\kappa_{1}N_{4}m_{3}+\kappa_{2}N_{2}m_{2}+\kappa_{2}N_{3}m_{3}-\frac{iN_{6}}{2}\kappa_{1}, \\
\partial_{t}m_{3} & =i\frac{\omega_{B}}{2}m_{3}-\kappa_{1}N_{2}m_{1}+\kappa_{1}N_{3}m_{1}+\kappa_{1}N_{4}m_{2}+\kappa_{1}N_{4}m_{3}+\kappa_{2}N_{2}m_{3}+\kappa_{2}N_{3}m_{2}-\frac{iN_{6}}{2}\kappa_{1}, \\
\partial_{t}m_{4} & =i\frac{\omega_{A}}{2}m_{4}+\kappa_{1}N_{1}m_{4}+\kappa_{1}N_{4}m_{1}-\kappa_{2}N_{1}m_{2}+\kappa_{2}N_{3}m_{1}+\kappa_{2}N_{3}m_{4}+\kappa_{2}N_{4}m_{2}-\frac{iN_{6}}{2}\kappa_{2}, \\
\partial_{t}m_{5} & =i(\omega_{A}+\omega_{B})m_{5}+\kappa_{1}N_{1}m_{5}+\kappa_{1}N_{4}m_{5}+\kappa_{1}N_{5}\left(m_{1}-m_{4}\right)+\kappa_{2}N_{2}m_{5}+\kappa_{2}N_{3}m_{5}+\kappa_{2}N_{5}(m_{2}-m_{3}), \\
\partial_{t}m_{6} & =i(\omega_{A}+\omega_{B})m_{6}+\kappa_{1}N_{1}m_{6}+\kappa_{1}N_{4}m_{6}+\kappa_{1}N_{6}\left(m_{1}-m_{4}\right)+\kappa_{2}N_{2}m_{6}+\kappa_{2}N_{3}m_{6}+\kappa_{2}N_{6}(m_{2}-m_{3}),
\end{align*}
\end{widetext}
where $M_j(t)$ and $N_j(t)$ are the coefficient functions after integration, in the operator $\bar{O}_y$ and $\bar{O}_z$ respectively.  $M_{j}(t)=\int_{0}^{t}{\rm d}s\beta(t,s)m_{j}(t,s)$ and $N_{j}(t)=\int_{0}^{t}{\rm d}s\alpha(t,s)n_{j}(t,s) (j=1,2,3,4)$. $N_{5(6)}(t,s')=i\int_{0}^{t}{\rm d}s'\alpha(t,s)n_{5(6)}(t,s,s')$, $M_{5(6)}(t,s')=i\int_{0}^{t}{\rm d}s'\beta(t,s)m_{5(6)}(t,s,s')$.
When $s' = t$, the boundary conditions are
\begin{align*}
n_{5}(t,s,t) & =-2i(\kappa_{1}n_{3}+\kappa_{2}n_{4})-iM_{5} \\
 & -2(M_{1}n_{3}+M_{2}n_{4}-M_{3}n_{1}-M_{4}n_{2}), \\
n_{6}(t,s,t) & =-iN_{5}, \\
m_{5}(t,s,t) & =-2i(\kappa_{1}m_{3}+\kappa_{2}m_{4})-iM_{6},\\
m_{6}(t,s,t) & =-iN_{6}-2(N_{1}m_{3}+N_{2}m_{4}-N_{3}m_{1}-N_{4}m_{2}).
\end{align*}
When $s= t$, the initial conditions are
\begin{align*}
&n_{1}(t,t)=\kappa_{1}-iM_{1}(t), \\
&n_{2}(t,t)=\kappa_{2}-iM_{2}(t), \\
&n_{3}(t,t)=-iM_{3}(t),\\
&n_{4}(t,t)=-iM_{4}(t), \\
&n_{5}(t,t,s')=-iM_{5}(t,s'),\\
&n_{6}(t,t,s')=-iM_{6}(t,s'), \\
&m_{1}(t,t)=-iN_{1}(t),\\
&m_{2}(t,t)=-iN_{2}(t), \\
&m_{3}(t,t)=-iN_{3}(t),\\
&m_{4}(t,t)=-iN_{4}(t),\\
&m_{5}(t,t,s')=-iN_{5}(t,s'),\\
&m_{6}(t,t,s')=-iN_{6}(t,s').
\end{align*}
With the determined coefficient functions, the reduced density matrix of the embedded two-qubit system can be recovered by taking the ensemble average over two noises,
\begin{align*}
\rho_t &= \mathcal{M}_{z,y}[|\psi_t\rangle \langle \psi_t| ]  \\
 &= \int\frac{\rm{d}^2 z}{\pi}\int\frac{\rm{d}^2 y}{\pi} |\psi_t\rangle \langle \psi_t |.
\end{align*}
The above discussion is based on an arbitrary correlation function $\beta(t,s)$,  depending on the nature of the environment. The universal approach applies to a broad context of the related studies on the generalized JC model in the presence of a leaky cavity mode or a cavity network.

\section{Numerical results}

With the derived QSD equation for the leaky cavity model, we can now study how the quantum features of the embedded system is mediated by the two-layer-noise in the parameter space. For simplicity, the Ornstein-Uhlenbeck noise $\beta(t,s)=\frac{\Gamma\gamma}{2}e^{-\gamma|t-s|}$ is used in the numerical simulation, where $\Gamma$ is the memory time parameter and $1/\gamma$ is the memory time of the environment. When $\gamma \rightarrow \infty$, the environment has zero memory time and the system undergoes a Markovian behavior. Particularly, we focus on two parameters, the coupling strength $g$ between the two-qubit system and the cavity mode, and the memory time parameter $\gamma$ from the external environment (it can also characterize the coupling strength between the cavity and the photon counter or detector).
\begin{figure}[htbp]
\centering

\subfigure{
\begin{minipage}{8cm}
\centering
\includegraphics[scale=0.45]{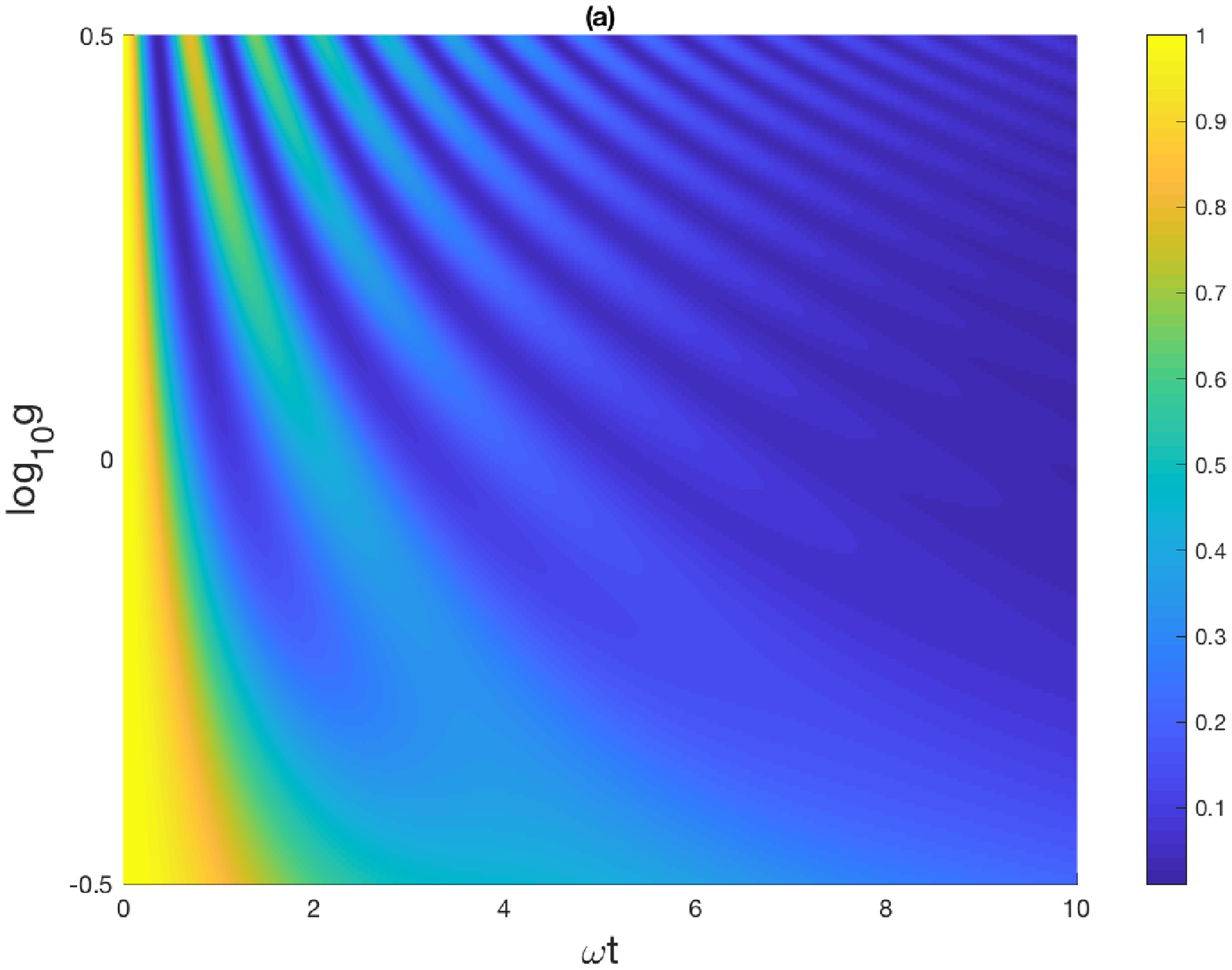}
\end{minipage}
}
\subfigure{
\begin{minipage}{8cm}
\includegraphics[scale=0.45]{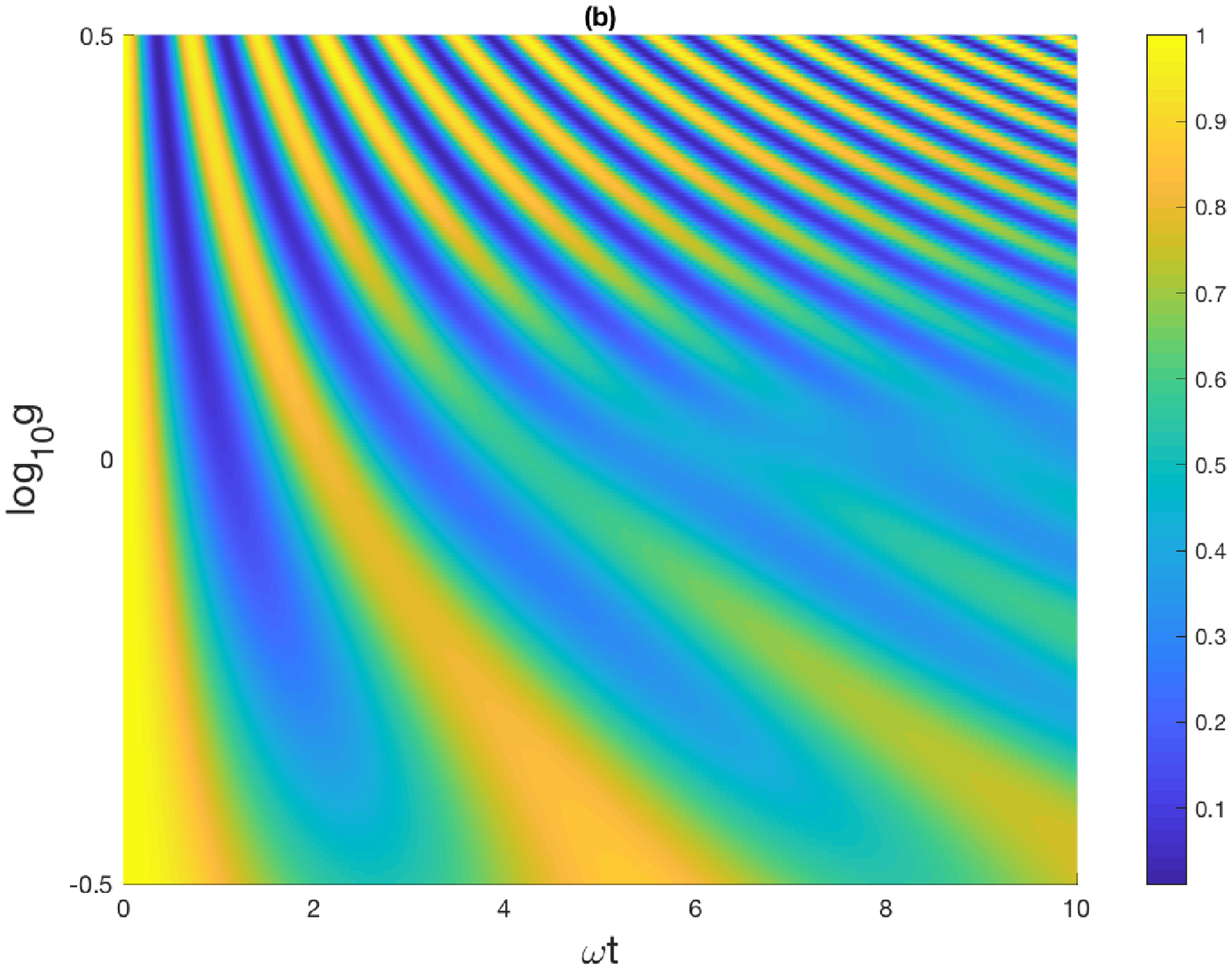}
\end{minipage}
}
\caption{Dynamics of entanglement in the two-qubit system for the varying coupling strength $g$. The initial state of the two-qubit system is the Bell state $|\Psi^+\rangle =(|10\rangle+|01\rangle)/\sqrt {2}$. The parameters are set as: $\omega_s = 2\omega$, $\kappa_1 = \kappa_2 =1$, $\Gamma=1$. In (a), the memory time parameter $\gamma= 5$. In (b), $\gamma =0.5$.}
\label{1001}
\end{figure}

The time evolution of the two-qubit entanglement is plotted in the Fig.~\ref{1001}. The two-qubit system is prepared maximal entangled in the Bell state $|\Psi^+\rangle =(|10\rangle +|01\rangle)/\sqrt{2}$. With different values of the memory time parameter $\gamma$, we study the dynamics of the entanglement in the Markovian (Fig.~\ref{1001} (a)) and non-Markovian (Fig.~\ref{1001} (b)) regimes respectively. When the environment is Makovian, the memory time parameter $\gamma =5$, the entanglement decreases from the beginning and then arises. This phenomenon indicates that the single cavity mode modifies the Markovian environment into a non-Markovian one. In Fig.~\ref{1001} (b), once we set the memory time parameter $\gamma = 0.5$, long-lasting and higher revived entanglement can be observed. However, we notice that when the coupling strength is in the range of $0.8<g<1.2$, the entanglement decays quickly even though the environment is in the non-Markovian regime. This phenomenon indicates that the back flow of information due to the non-Markovian environment is not always helpful for protecting the coherence of the system and its influence on the system has to be studied as a result of the competition between two mechanisms: (i) the interaction between the system and the cavity mode; (ii) the influence from the external environment.
\begin{figure}[htbp]
\centering
\subfigure{
\begin{minipage}{8cm}
\includegraphics[scale=0.45]{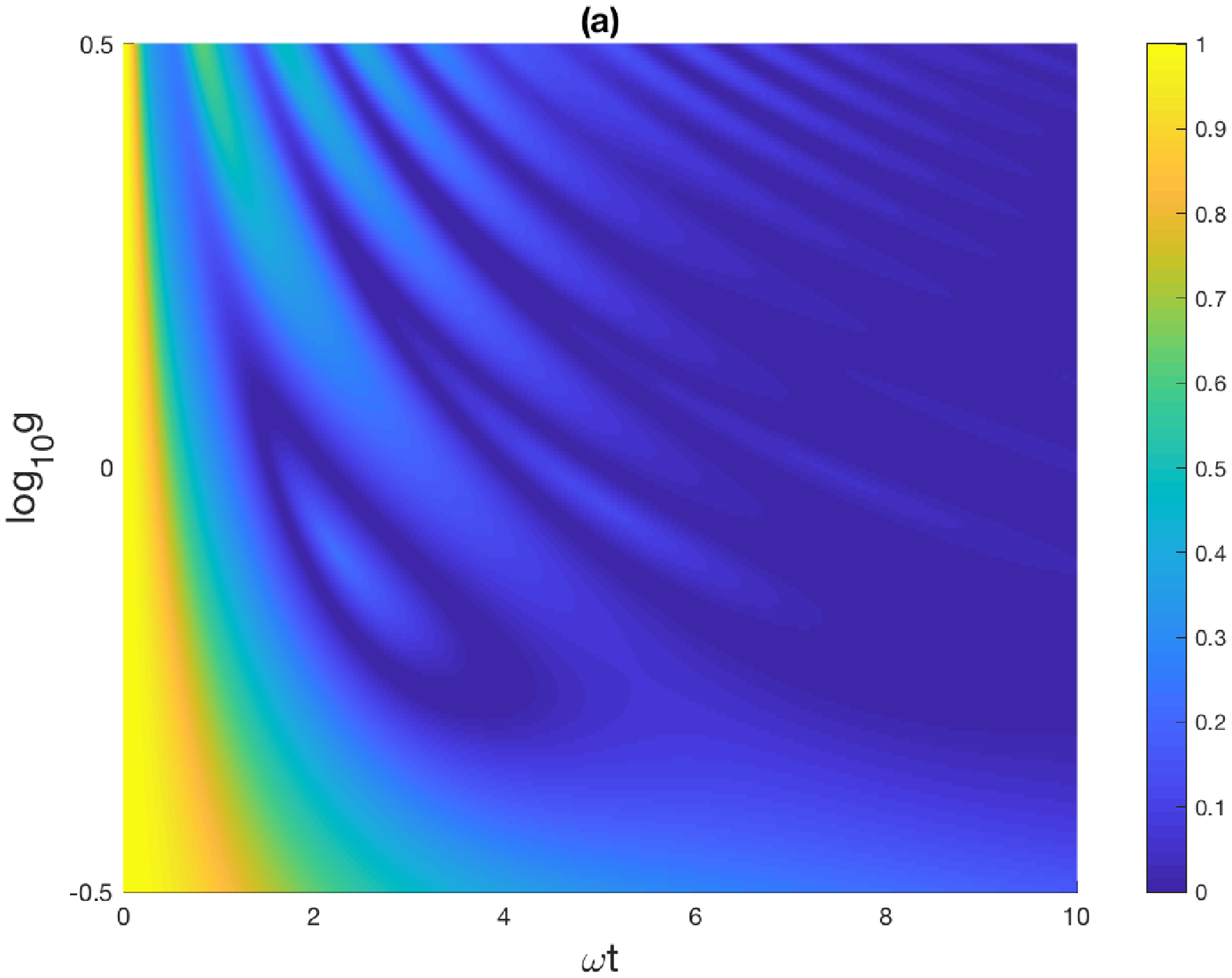}
\end{minipage}
}
\subfigure{
\begin{minipage}{8cm}
\includegraphics[scale=0.45]{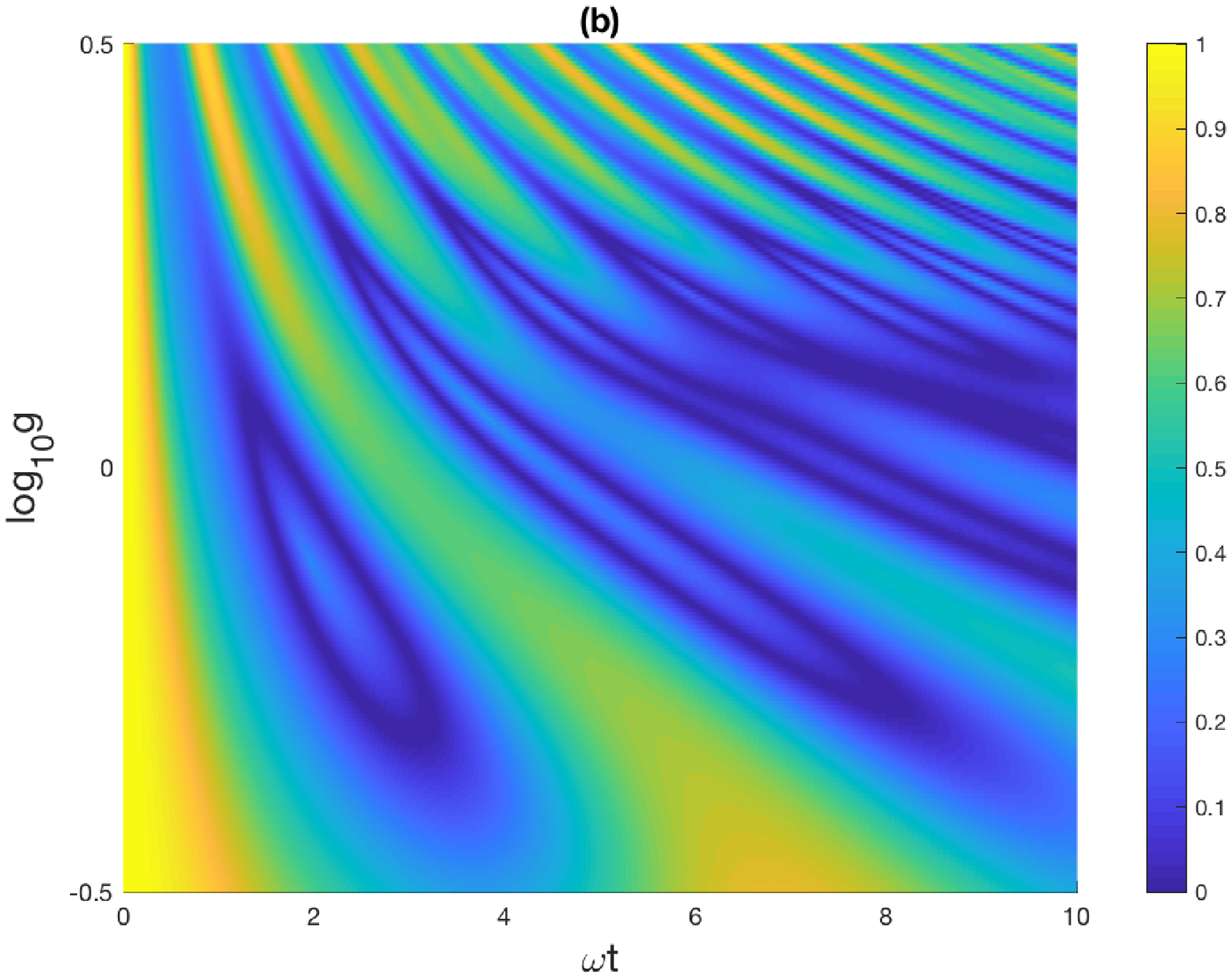}
\end{minipage}
}
\caption{Dynamics of entanglement in the two-qubit system for the varying coupling strength $g$. The initial state of the two-qubit system is the Bell state $|\Phi^+\rangle =(|11\rangle+|00\rangle)/\sqrt{2}$. The parameters are set as: $\omega_s = 2\omega$, $\kappa_1 = \kappa_2 =1$, $\Gamma=1$. In (a), the memory time parameter $\gamma= 5$. In (b), $\gamma =0.5$.}
\label{1100}
\end{figure}

In Fig.~\ref{1100}, the dynamics of the entanglement in the two-qubit system is simulated when the initial state is prepared as the Bell state $|\Phi^+\rangle = (|11\rangle +|00\rangle)/\sqrt{2}$. The dynamics of entanglement is modulated due to the change in the memory time parameter $\gamma$, but they hold similar fringe as in Figs.~\ref{1100} (a) and \ref{1100} (b). For this initial state, the non-Markovian environment speeds up the evolution of entanglement. In Fig.~\ref{1100} (b), when the coupling strength $g$ is larger than $3$, the coherent time becomes longer and the entanglement can be restored close to $1$, which means now the system is maximally entangled. This revival entanglement is induced by the interaction between the system and the cavity mode directly, so it occurs at a higher frequency. When the coupling factor $g$ is small, i.e., the coupling between system and cavity mode is weak, meanwhile the influence from the external environment is comparatively strong, so that it will take more time to observe the revival. Figure~\ref{1100} (b) clearly demonstrates two mechanisms affecting the dynamics of the embedded system, which turn out to be a trade-off problem.
\begin{figure}[htbp]
\centering
\subfigure{
\begin{minipage}{8cm}
\includegraphics[scale=0.6]{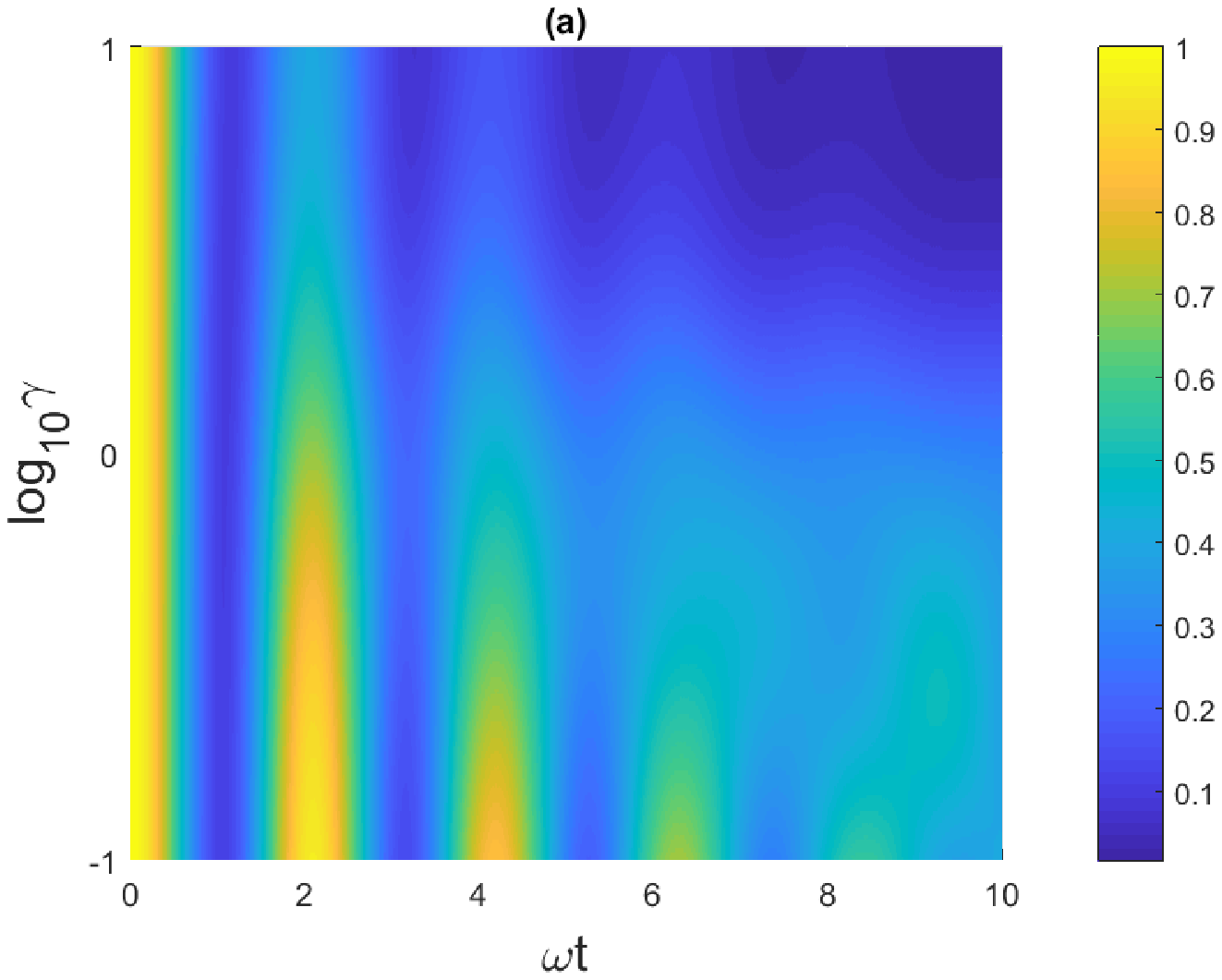}
\end{minipage}
}
\subfigure{
\begin{minipage}{8cm}
\includegraphics[scale=0.6]{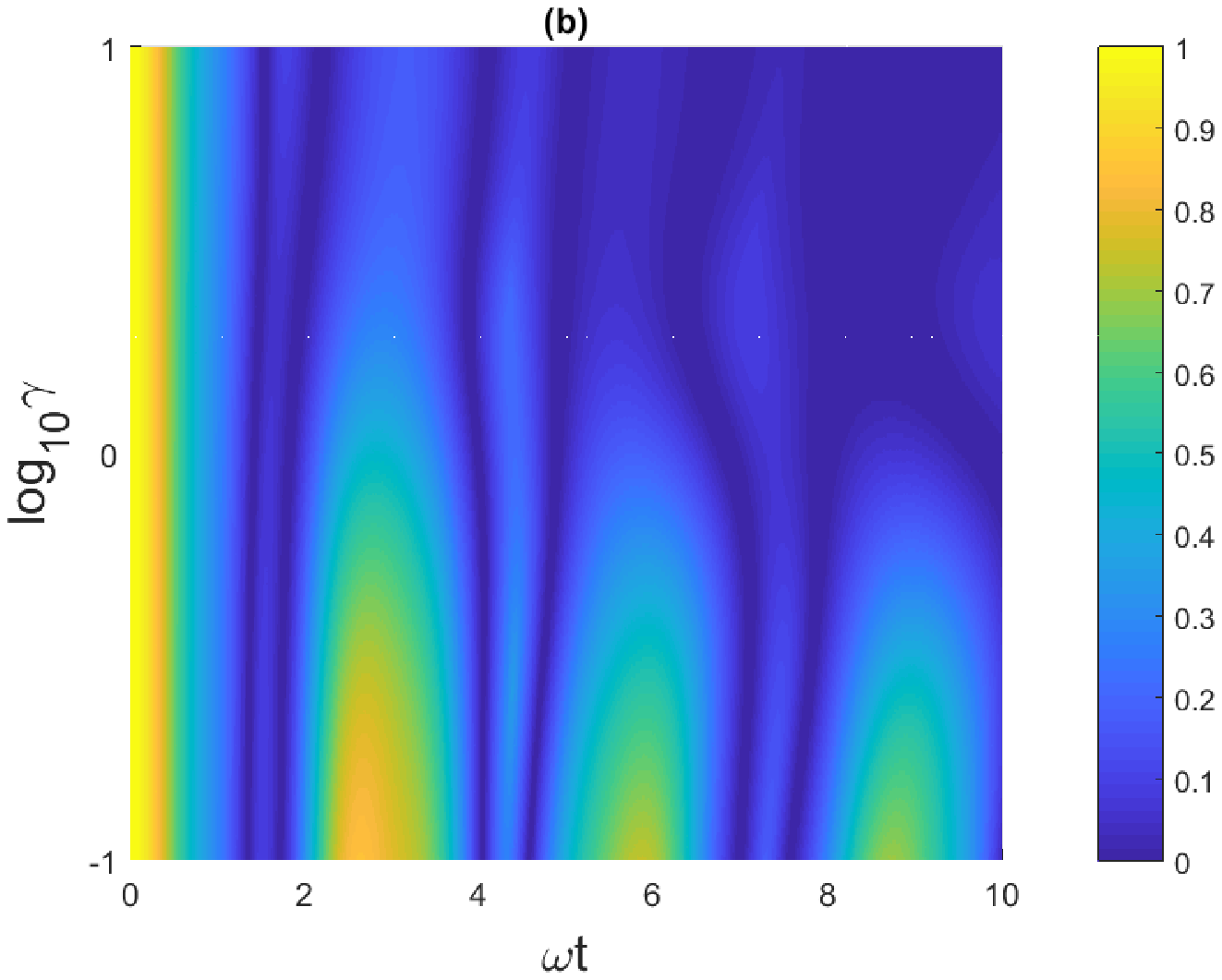}
\end{minipage}
}
\caption{Dynamics of entanglement in the two-qubit system for the varying coupling strength $g$. The initial state of the two-qubit system is a Bell state, (a) $|\Psi^+\rangle =(|10\rangle+|01\rangle)/\sqrt{2}$, (b) $|\Phi^+\rangle = (|11\rangle +|00\rangle)/\sqrt{2}$. The parameters are set as: $\omega_s = 2\omega$, $\kappa_1 = \kappa_2 =1$, $g=1$. }
\label{gamma}
\end{figure}

In Fig.~\ref{gamma}, we demonstrate the dynamics of entanglement with the varying memory time parameter $\gamma \in [0.1, 1]$. The initial state is prepared as the Bell state $|\Psi^+\rangle = (|10\rangle +|01)/\sqrt{2}$ and $|\Phi^+\rangle = (|11\rangle +|00\rangle)/\sqrt{2}$. The time-evolution behaviors for these two Bell states are similar to each other. The upper bound shows the dynamics when the environment is Markovian, where the entanglement decays in a nearly-exponential way. While on the bottom side, the entanglement lasts for a long time.

\begin{figure}[htbp]
\centering
\subfigure{
\begin{minipage}{8cm}
\includegraphics[scale=0.45]{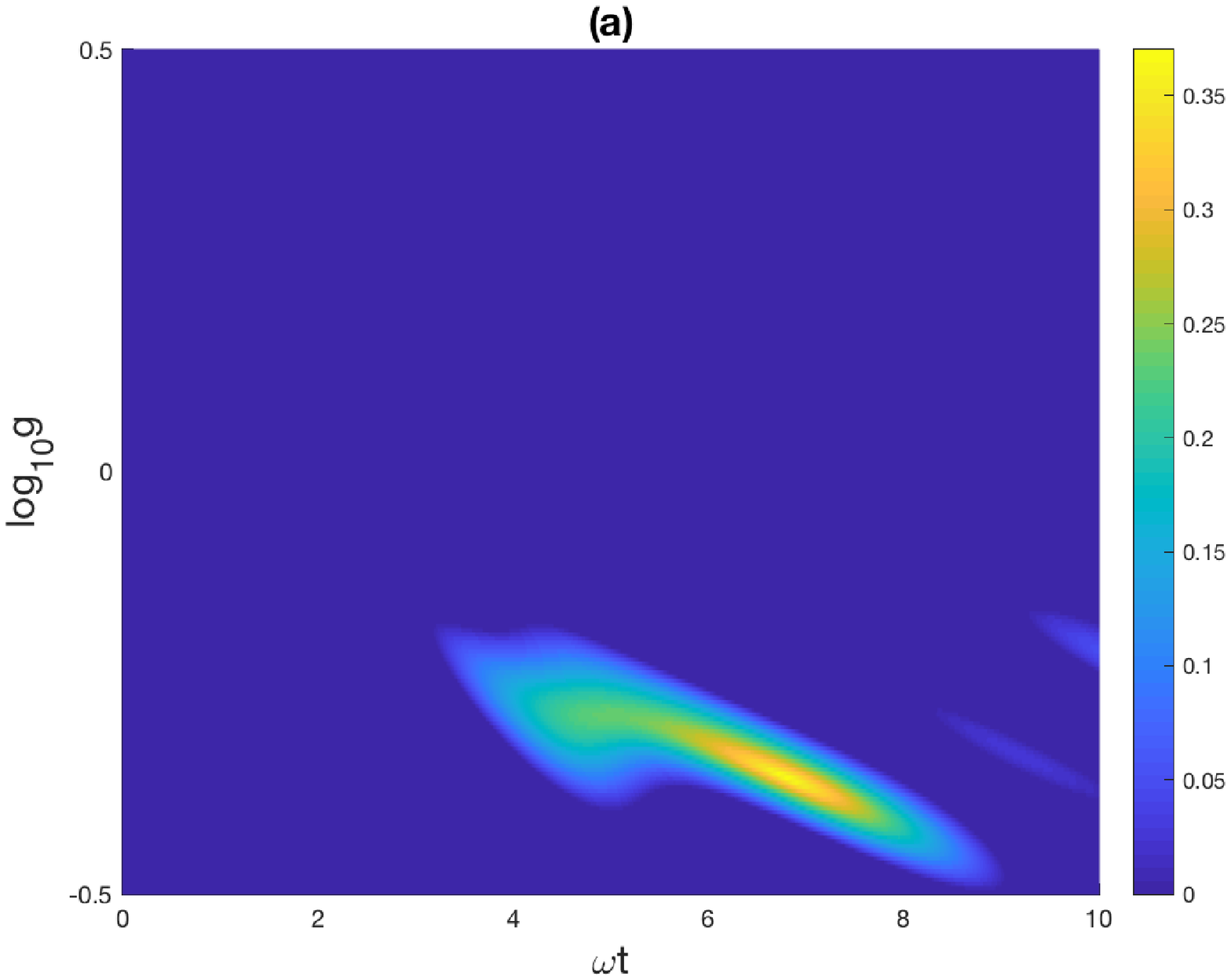}
\end{minipage}
}
\subfigure{
\begin{minipage}{8cm}
\includegraphics[scale=0.45]{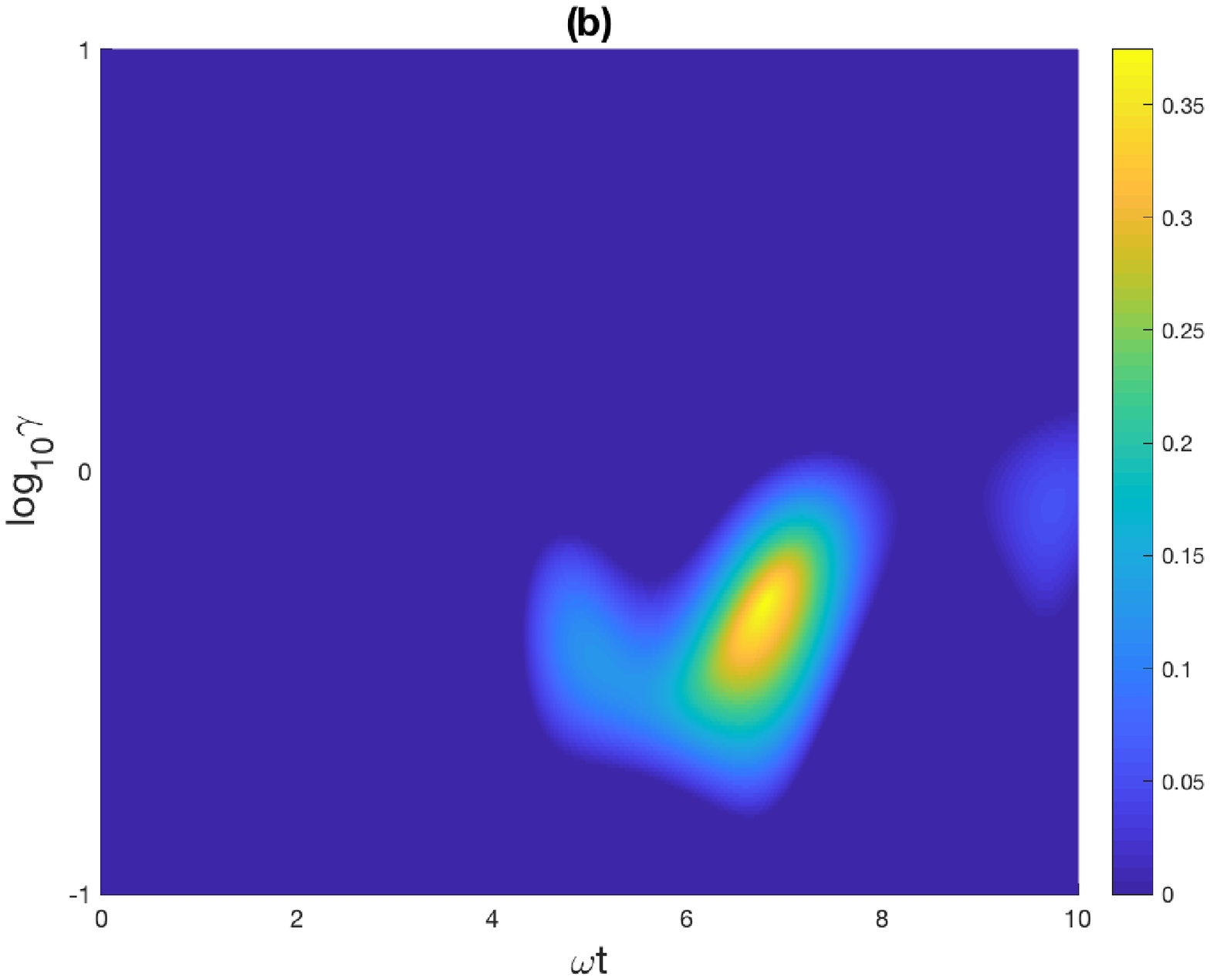}
\end{minipage}
}
\caption{Rebirth of entanglement in the two-qubit system. The initial state of the two-qubit system is $|11\rangle$. The parameters are set as: $\omega_s = \omega$, $\kappa_1 = \kappa_2 =1$. (a), the memory time parameter $\gamma =0.5$. (b), the coupling strength $g=0.4$. }
\label{reso}
\end{figure}

In addition, we simulate the dynamics starting with different initial states. Firstly, we set the initial state is $|11\rangle$. Secondly, we set the cavity is in resonant with the two qubits ($\omega_s = \omega $) and coupled with them symmetrically ($\kappa_1 = \kappa_2$). Numerical simulations show that there is no entanglement generation when there is no external environment. However, the entanglement generation is observed once the system is coupled to the non-Markovian environement, as shown in the Fig.~\ref{reso}.

We notice that the entanglement generation only happens approximately in an overlap area that $0.4 < \gamma< 0.6$ and $0.15<g<0.8$. In Fig.~\ref{reso} (a), we fix the $\gamma = 0.5$ and investigate the dynamics of entanglement by varying $g$. When $g$ is large enough, the strong coupling approximately isolates the system from the external environment, so that there is no entanglement generation. On the contrary, when $g$ is small, no entanglement generation is observed because the two-qubit system is disconnected to the cavity and environment. Only when the coupling strength is set $g \approx 0.4$, the concurrence approaches to the maximium.

An intuitive understanding is that a longer memory time of the environment allows one to retrieve more information from the environment, which is helpful to strengthen the generation of quantum coherence and entanglement in the system. However, our numerical simulations show the counter-intuitive results as in Fig.~\ref{reso} (b) (setting $g=0.4$). When $\gamma >0.8 $, the environment is approaching the Markov limit, and the two qubits keep separated all the time. The entanglement generation happens when $\gamma$ decreases to $0.5$ and the environment is typically in the non-Markovian regime. And our numerical results show that the strength of entanglement generation is not monotonically increased when the environment's memory time is further increased.  The system entanglement disappears when $\gamma$ exceeds the lower boundary $0.4$.

\section{Conclusion}
In conclusion, we study a double-qubit JC model in a leaky cavity mode. In reality, the cavity is not connected to a Markovian environment, but a non-Markovian one. In order to understand the dynamics of the embedded system and how it is influenced by the cavity together with the external environment, we derive the exact QSD equation for the embedded system. Technically we have to deal with a double-layer environment for this open system. We follow the principle idea of QSD approach and expand the trajectories of the system in the basis of two independent noises.

Another major conclusion is that the dynamics of the system is modulated by two mechanisms. The competition between the non-Markovian features from the environment and the direct coupling to the cavity shapes the evolution of the system. Counter-intuitively, the coupling strength $g$ and the memory time parameter $\gamma$ jointly determine the entanglement generation. This sheds light on understanding the complexity of non-Markovian dynamics. Also it helps optimize the parameters to enhance the coherence time and entanglement in a measurement scheme by a readout detector. Our work paves the road to obtain the exact dynamics of the general JC model in the presence of the leakage cavity or cavity network.

\end{document}